\documentclass{article}

\usepackage{amsmath,amssymb,amsfonts,epsfig, float, setspace, lineno}

\newcommand{\poff} {p_{\text{\emph{off}}}}
\newcommand{\pkill} {p_{\text{\emph{kill}}}}

\newcommand{\pmut} {p_{m}}

\psfigdriver{dvips}

\newcommand{\bra}[1]{\langle #1|}
\newcommand{\ket}[1]{|#1\rangle}

\renewcommand\thefootnote{\fnsymbol{footnote}}

\begin{document}

\title{Understanding clustering in type space using field theoretic techniques}

\author{{\bf Daniel John Lawson\footnotemark[1]}\\
\emph{Biomathematics and Statistics Scotland,}\\ \emph{Macaulay Institute, Aberdeen, United Kingdom, AB15 8QH.}\\
\\
{\bf Henrik Jeldtoft Jensen\footnotemark[2]}\\
\emph{Department of Mathematics, Imperial College London,} \\ \emph{South Kensington campus, London, United Kingdom, SW7 2AZ.}
}


\date{\today}

\maketitle

\doublespacing

\begin{abstract}
The birth/death process with mutation describes the evolution of a population, and displays rich dynamics including clustering and fluctuations.  We discuss an analytical `field-theoretical' approach to the birth/death process, using a simple dimensional analysis argument to describe evolution as a `Super-Brownian Motion' in the infinite population limit.  The field theory technique provides corrections to this for large but finite population, and an exact description at arbitrary population size.  This allows a characterisation of the difference between the evolution of a phenotype, for which strong local clustering is observed, and a genotype for which distributions are more dispersed.  We describe the approach with sufficient detail for non-specialists.
\fnsymbol{footnote}
\footnotetext[1]{Electronic address: daniel@bioss.ac.uk}
\footnotetext[2]{Electronic address: h.jensen@imperial.ac.uk}
\renewcommand\thefootnote{\arabic{footnote}}
\end{abstract}


\linenumbers
\section{Introduction}

Throughout the biological literature, the term ``diffusion in genotype space'' is used to describe a population acting under genetic drift in the absence of selection.  This is not diffusion at the \emph{individual} level, but at the \emph{population} level, where the individuals form clusters resembling a species, the mean position of which performs a random walk i.e. diffuses.  The `species' may consist of a number of clusters at any given time. However, these clusters remain close together, and the species is limited in the maximal width that it can achieve in any given direction in the space \cite{Lawson06-Neutral}.

Clustering phenomena are well understood for reproducing and dying organisms dispersing in real space \cite{ZhangEtAl90-Diffusion,Meyer96-clustering}.  In the case of real space, the relationship of the microscopic process to the stochastic Partial Differential Equation (PDE) formalism is clear, due to the (exact) field theory mapping \cite{TauberHowardLee05} of the underlying microscopic process to a stochastic PDE.
However, no such translation has been done in the case of type (sequence) space, be it of a genotype or a phenotype, where clustering phenomena are also observed \cite{Lawson06-Neutral,DerridaPeliti91,DerridaEtal99}.  We perform the translation and show that reproducing and dying organisms either diffusing in real space or mutating in type space are fundamentally the same process in an infinite population.  This equivalence only applies in the infinite population limit, and so we provide finite size corrections to the stochastic PDE, allowing for individuals to mutate only on a birth event.

From known results for diffusing organisms, there exists an `upper-critical dimension' $d_c$, above which general `mean field' results hold but below which the behaviour is different.  
A phenotype forms a one dimensional type space, which can be thought of as a single trait.  Conversely, we consider a genotype as a very long amino acid string, and hence genotype space is high dimensional as mutations are free to occur at a large number of independent positions.
 Therefore there exists an important distinction between the evolution of a given phenotype, and a genotype.  The theory of Critical Branching Processes \cite{Slade02-SuperBrownian} finds that in high dimensions describing genotype space ($d > d_c$, where in our case the critical dimension $d_c = 2$  \cite{Winter02-Branching}), birth/death dynamics are described fully by the lineages. A lineage remains distinct until all individuals in it die. However, in low dimensions ($d \leq d_c$) describing a particular phenotype, additional clustering within the distribution of the lineage occurs. Although sometimes distinct, the clusters in phenotype space can merge, and hence clusters are poorly defined entities. Instead, a careful average over the distribution called a `peak' provides a more useful description of phenotypes \cite{Lawson06-Neutral}.

Critical Branching Processes have a total population that does a random walk and only surviving lineages with $N(t_{final}) > 0$ are considered. For real space birth/death processes, the same phenomological clustering and upper-critical dimension $d_c = 2$ are also found when considering systems of fixed population size \cite{ClusteringNeutral}. As we show that neutral evolution has the same description in the infinite population limit as Critical Branching Processes in real space, this result also applies to evolution, and for qualitative studies we can choose whether to consider systems of fixed or fluctuating total population.


We use the technique of second quantisation of a master equation and mapping to a field theory\cite{TauberHowardLee05}, for which most previous work focuses on the infinite population limit.  However, field theory is a good tool for obtaining analytical results for finite and changing population sizes, as is the case for real populations.  The technique was developed in the setting of reaction-diffusion systems, where particles diffuse continuously, unlike our case of diffusion in a type space \emph{where the diffusion only occurs on a birth or death event}.

This paper addresses three main issues:
\begin{enumerate}
\item We discuss how a microscopic model of evolution can be represented as a Field Theory, and derive the stochastic PDE that follows in this case.
\item Understanding the `asymptotic' behaviour, i.e. the infinite population and long wavelength properties of evolution, by identifying that this is a solved problem.  Super Brownian Motion and Critical Branching Processes provide a wealth of results which are made explicit by the comparison of the relevant stochastic PDEs.
\item Relating the asymptotic behaviour to the behaviour at finite population size.
\end{enumerate}

\subsection{Known results for clustering due to birth/death processes}
\label{subsec:knownres}

We will use known results of birth/death processes from the field of Critical Branching Processes, which tackles similar problems to field theory techniques but differs in approach and terminology.  Refs. \cite{Slade02-SuperBrownian} and  \cite{Winter02-Branching} give more details, and a more direct technique is used in \cite{ClusteringNeutral}.


As discussed above, the behaviour is qualitatively different above and below a critical dimension $d_c$.  In the language of statistics, for $d \leq d_c$ the only stable solution is a $\delta_0$ measure in the correlations between individuals - i.e. all individuals are fully correlated in their position.  This implies that all individuals are localised in space so that their distribution collapses to a point when viewed at very large length-scales.  In unscaled type space, this corresponds to a local peak \emph{with a characteristic width} ($s$) i.e. a length-scale.  The length-scale $s$ scales to $0$ in the infinite population limit for $d \leq d_c$.  However, for $d>d_c$ other non-trivial measures exist describing the correlations in the system, which correspond to a distribution of individuals spread out over a number of lineages.  The distribution of time since last ancestor forms a power-law distribution in the infinite population limit, and since lineages typically do not intersect in high dimensions, there is no characteristic width to the distribution (and hence no length-scale).  

The theorems available for the clustering process are usually devoted to deriving general behavioural properties, such as the convergence to either of the above measures in various dimensions.  Many clustering phenomena are described by the same scaling relations for $d > d_c^i$, where $i$ is a label for a particular phenomena (e.g. birth/death processes in Euclidean space, as our model).  Thus each model may have a different critical dimension, but above that the scaling behaviour is the same.  Examples are given in \cite{Slade02-SuperBrownian}: Galton Watson Trees embedded into space (which is the real space diffusion version of our model) have $d_c=2$, Directed Percolation \cite{TauberHowardLee05} has $d_c=4$, Percolation has $d_c=6$ and Lattice Trees (a lineage tree embedded in a lattice so that separate lineages never meet) have $d_c=8$.  All dimensions refer to the number of spatial dimensions - the stochastic process consists of the extra dimension of time.  Below the critical dimension each model behaves specially, however above the critical dimension all models follow the same scaling relation.  All of these models can be described as a birth/death process embedded in some type of space.

Super Brownian Motion is the limiting process of all of the above processes for $d > d_c$.  This can be described by a stochastic PDE in many cases.  In the case of Galton-Watson Trees \cite{CoxKlenke03}, in terms of the density $\rho$ as a function of space $x$ and time $t$, the stochastic PDE is:

\begin{equation}
\frac{\partial \rho(\mathbf{x},t)}{\partial t} = D \bigtriangledown^2 \rho(\mathbf{x},t)  + c \sqrt{\rho(\mathbf{x},t)}\eta(\mathbf{x},t), \label{eqn:GWtree}
\end{equation}
where $D$ is the diffusion constant and $c$ is a constant describing the magnitude of the noise.  We will obtain this functional form in the infinite population limit for the case of evolution in type space, which with some dimensional analysis means that evolution as we define it has $d_c=2$.

\subsection{The model}

We consider $N(t)$ individuals at time $t$, and each individual has a discrete type $x \in \mathcal{Z}^d$, which it \emph{retains during its lifetime}.  The dimension of type space $d$ is arbitrary in the formalism.  A timestep consists of performing a birth attempt with probability $\poff/(\poff+\pkill)$, or otherwise a killing attempt occurs\footnote{This is an example of the Gillespie algorithm \cite{Gillespie76}.}.  We focus on the case $\pkill=\poff$ throughout this discussion.  Time is measured in generations and increases by the average waiting time between events, $1/N(\poff+\pkill)$ per timestep.

\begin{itemize}
\item Birth attempt:  A parent individual (with type $x$) is selected at random, and an offspring is created with type $x$ and mutated with probability $\pmut$.  A mutation involves $x$ changing by $\pm 1$ in a randomly chosen direction.
\item Killing attempt: A randomly chosen individual is removed.
\end{itemize}

\begin{figure}
[htb]
\begin{center}
\epsfig{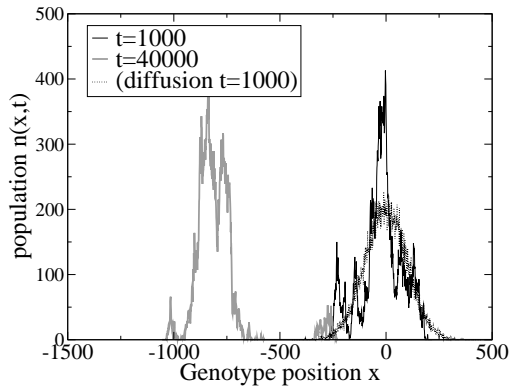}
\caption{Sample distribution for $N=10000$ individuals starting at $0$ and evolving in a 1-dimensional type space.  For early time, (black line) the distribution behaves similarly to diffusion, but once the peak has become `large', it begins to move around, and can split up into a number of clusters (as shown by grey line).  The  distribution for $N=10000$ diffusing particles \cite{Lawson06-Neutral} is also shown (a normal like distribution centred at zero, dotted line) at early time; the width increases as $\sqrt{t}$.  \label{fig:sampledistrib}}
\end{center}
\end{figure}

This definition differs from the case of a birth/death process with diffusion in real space, where all individuals are diffusing constantly between birth/death events.  Our model permits `diffusion' only on birth events via mutation.  A sample run is shown in Figure (\ref{fig:sampledistrib}) for an early and late time, and a sample distribution from diffusion is shown for comparison.  Simulations at different $N$ show that there is a clustering behaviour that persists regardless of $N$ (not shown, though see \cite{Lawson06-Neutral}).  We wish to understand how the clustering depends on dimensionality, both qualitatively and quantitatively.

\section{Field theory approach to obtain a Stochastic PDE}
Doi's process of second quantisation \cite{Doi76} is used to obtain a Field Theory from a Master Equation. 
A detailed description of the mapping process is presented in \cite{TauberHowardLee05}, and a detailed background can be obtained from \cite{LeBellacFT}.  

\subsection{Outline of the method}
\begin{sloppypar}We outline the method for obtaining a field theory from a Master Equation of the form $dP(\{n\},t)/dt=f(\{n\})$.  Here $P(\{n\},t)$ is the probability distribution of the state $\{n\}= \{n_1,\cdots,n_i,\cdots,n_{L}\}$, where $L$ is the size of the type space and $n_i$ is the population size of a given type $i$.\end{sloppypar}
\begin{enumerate}
\item Define the state $\ket{\{n\}}$, and use the equation for $dP(\{n\},t)/dt$ to obtain an equation of motion for $\ket{\{n\}}$.
\item Define $\ket{\{\phi\}}$ as the superposition of all possible states with probabilities $P(\{n\},t)$. This provides an equation of motion of the form $d\ket{\{\phi\}}/dt = -\hat{H}\ket{\{\phi\}}$, where $\hat{H}$ is called the \emph{Hamiltonian}.
\item The state $\ket{\{\phi\}}$ is expressed in terms of operators; the field $\phi(\mathbf{x},t)$ emerges as a natural quantity in the system, being the eigenvalue of the so called `creation' operator, which counts the number of individuals and hence is related to the density.  The field $\phi(\mathbf{x},t)$ is simply a complex number defined for all spatial points $x$.
\item By taking the continuous space limit, the equations for $\phi(\mathbf{x},t)$ become tractable.
\item Observables $\overline{A}$ can be related to $\phi$ in terms of functional integrals.  However, most quantities of interest such as the density can be obtained directly from examination of the Hamiltonian alone.
\item A stochastic PDE in our case can easily be obtained from $H$, providing access to other techniques or, in our case, previous results.
\end{enumerate}


We find that additional terms appear in the field theory due to the insistence that movement only occurs on a birth; these are difficult to deal with in analytical calculations.  These terms are negligible in the infinite population limit, in which case our model reduces to the real space birth/death process with diffusion of individuals.  This is a previously solved case, as discussed in Section \ref{subsec:knownres}.  We are also concerned with the finite population case, for which we provide both an exact and an approximate stochastic PDE.

\subsection{To Second Quantised form}
We now follow the procedure of second quantisation of a Master Equation, providing explicit details only for the one-dimensional case for readability.  The starting point is the master equation for the Evolution process as defined above. The equation for the change in the probability distribution $p(\{n\})$ over all sites $\{n\} = (n_{1},\cdots,n_i,\cdots,n_{L-1},n_L) \equiv n_i$, in notation enumerating only sites different from $\{n\}$ for brevity, is:
\begin{align}
\Delta_t p(n_i;t)= \nonumber \\
\frac{\poff}{2N}\sum_{i}\Big\{&n_i [\pmut p(n_{i-1}-1;t) + \pmut p(n_{i+1}-1;t) - 2 p(n_i;t)] \nonumber\\
& + 2(1-\pmut )(n_i-1) p(n_i-1;t) \nonumber\\
&+ 2(n_i+1)p(n_i+1;t) -2n_i p(n_i;t)\Big\}. \label{eqn:MEevol}
\end{align}
Eq. (\ref{eqn:MEevol}) follows directly from a microscopic description of the model.  We sum over all possible lattice points $i$ where a change could occur.  The terms from left to right are, on the top line: enter state $\{n\}$ by a birth at $i$ mutating left; mutating right; leaving state $\{n\}$ by a birth at $i$. Second line: entering state $\{n\}$ by a birth without mutation.  Third line: entering state $\{n\}$ by a death at $i$, and leaving state $\{n\}$ by a death at $i$.  We have ignored boundary terms as we will take $L \to \infty$.

The state $\ket{n_i}$ of a lattice point $i$ is defined as the number of individuals on it.  
We then define the state of the system $\ket{\{n\}}=\ket{n_1}\otimes\cdots\otimes\ket{n_L}$, where $\otimes$ denotes the outer product.

Eq. (\ref{eqn:MEevol}) is multiplied by $\ket{\{n\}}$, and we then relabel the states within the sum to ensure that the probabilities in all terms are expressed in terms of $p(\{n\})$, allowing the `state' vector to become different from $\ket{\{n\}}$.  Then we can define operators acting on the state in order to recover all terms in the summed state $\ket{\{n\}}$.  These operators also capture multiplicative terms in the number of individuals $n_i$.  We define the operators, called the annihilation operator $\hat{a}$ and the creation operator $\hat{a}^\dagger$, by their commutation relations:

\begin{align}
[\hat{a}_i,\hat{a}^\dagger_j] = \delta_{ij}, \\
[\hat{a}_i,\hat{a}_j] = [\hat{a}^\dagger_i,\hat{a}^\dagger_j] = 0.
\end{align}
The notation $[\hat{a}_i,\hat{a}^\dagger_j]$ means simply $\hat{a}_i \hat{a}^\dagger_j -  \hat{a}^\dagger_j \hat{a}_i$.  If we define the `vacuum lattice' $\ket{0}$ by $\hat{a}_i \ket{0} = 0$ for all $i$, and $\ket{n_i} = (\hat{a}^\dagger_i)^{n_i}\ket{0}$ then it is simple to show that the operators follow:

\begin{align}
\hat{a}_i\ket{n_i} = n_i \ket{n_i-1},\label{eqn:aini}\\
\hat{a}^\dagger_i\ket{n_i} = \ket{n_i+1}.\label{eqn:aidaggerni}
\end{align}
On multiplication of Eq. (\ref{eqn:MEevol}) by the state $\ket{\{n\}}$, summation over all states $\{n\}$, performing the relabelling and using the creation and annihilation operators, we find:

\begin{align}
\Delta_t \sum_{\{n\}} p(\{n\};t)\ket{\{n\}} = \frac{\poff}{2N}\sum_{\{n\}} \sum_i & p(\{n\};t) \Big\{\pmut  \hat{a}^\dagger_{i-1} \hat{a}^\dagger_{i} \hat{a}_{i} + \pmut  \hat{a}^\dagger_{i-1} \hat{a}^\dagger_{i} \hat{a}_{i} - 2 \hat{a}^\dagger_{i} \hat{a}_{i}  \nonumber\\
&+ 2(1-\pmut )(\hat{a}^\dagger_{i})^2 \hat{a}_{i} +2 \hat{a}_{i} - 2 \hat{a}^\dagger_{i} \hat{a}_{i}\Big\} \ket{\{n\}},
\end{align}
Or we can write this in (quasi)Hamiltonian form, using the notation $\ket{\{\phi\}}=\sum_{\{n\}}p(\{n\};t)\ket{\{n\}}$:

\begin{equation}
\Delta_t \ket{\{\phi\}} =-\hat{H} \ket{\{\phi\}},
\end{equation}
with the Hamiltonian:

\begin{equation}
\hat{H} = 
\frac{\poff}{2N} \sum_i \left[-\pmut  (a_{i-1}^\dagger+a_{i+1}^\dagger-2 a_{i}^\dagger) a_i^\dagger a_i - 2 (a_i^\dagger-1)^2 a_i \right]. \label{eqn:SQevol}
\end{equation}
This completes the mapping to second-quantised form.

\subsection{From Second Quantisation to Field Theory} \label{subsec:tofieldtheory}

The next step involves constructing so called `coherent states' such that $\hat{a}_{i}\ket{\phi} = \phi_i\ket{\phi}$, and  $\bra{\phi}\hat{a}^\dagger_{i} = \bra{\phi} \phi^*_i$ as described in \cite{TauberHowardLee05}.  The eigenvalues $\phi_i$ and $\phi^*_i$ of the $\hat{a}$ and $\hat{a}^\dagger$ operators respectively are complex numbers at a given point, and therefore in the continuous limit form a \emph{field} $\phi$ in space $x$.  This allows one to calculate observables by use of a projection state.  Here we simply use the results that $\hat{a}_i \rightarrow \phi_i$, $\hat{a}^\dagger_i \rightarrow \phi^*_i$.  
The continuous limit is then taken by allowing $\sum_i \rightarrow \int h^{-1} dx$, $\phi_i \rightarrow \phi(\mathbf{x},t) h$ and $\phi_i^* \rightarrow \tilde{\phi}(\mathbf{x},t)$, where we take $h$ (the distance between lattice points) to zero.  It can be shown that $\langle \phi \rangle = \langle n(\mathbf{x},t)/N \rangle$.  We consider the $\tilde{\phi}(\mathbf{x},t)$ and $\phi(\mathbf{x},t)$ fields to be independent.  This completes the mapping to a field theory, in terms of the \emph{Action} in the Statistical Weight:

\begin{equation}
S[\tilde{\phi},\phi] = \int d^dx\left\{-\phi(t_f)-\tilde{\phi}(0)[1-\overline{n}_0] + \int_0^{t_f}\left[\tilde{\phi}\partial_t\phi + H(\phi,\tilde{\phi}) \right]dt \right\},\label{eqn:action}
\end{equation}
expressed in terms of final time $t_f$, and the average initial occupancy $\overline{n}_0$.  The Action is related to the expectation of an observable $A$ by:

\begin{equation}
\overline{A}(t) = \mathcal{N}^{-1} \int\left( \lim_{L \to \infty} \prod_{i=1}^L\mathcal{D}\phi_i\mathcal{D}\phi^*_i\right)
A(\{\phi\}_t) \exp[-S(\{\phi^*\},\{\phi\})_0^t]. \label{eqn:obsexp}
\end{equation}
We have introduced a normalisation factor $\mathcal{N}$ and the path integral notation $\mathcal{D}\phi_i$ \cite{TauberHowardLee05}.
 Path integrals of the form of Eq. (\ref{eqn:obsexp}) have been well studied and we will discuss some of methods available to avoid performing explicit integration, by considering the Action $S$ directly.  The Action depends only on the Hamiltonian which we have derived from the Master Equation above.
Following this process for our case of neutral evolution from Eq. (\ref{eqn:SQevol}) we have:

\begin{equation}
H(\mathbf{\phi},\mathbf{\phi^*}) = \frac{\poff}{2N} \sum_i \left[-\pmut  (\phi_{i-1}^*+\phi_{i+1}^*-2 \phi_{i}^*)\phi_{i}^* \phi_i  - 2 (\phi_i^*-1)^2 \phi_i \right],
\end{equation}
or in the continuum limit:

\begin{equation}
H(\phi,\tilde{\phi}) = D \int d^dx \left[-(\bigtriangledown^2 \tilde{\phi}(\mathbf{x},t)) \tilde{\phi}(\mathbf{x},t)\phi(\mathbf{x},t)  - \frac{2}{\pmut } (\tilde{\phi}(\mathbf{x},t)-1)^2 \phi(\mathbf{x},t) \right]. \label{eqn:Hunshifted}
\end{equation}
We have introduced the Diffusion Constant $D=(\poff\pmut/2)(h^2/Ndt)$, which is kept constant when taking the limit.  The distance between types is $h$ and $dt$ is the timestep.  This equation is recovered for any dimensionality of Eq. (\ref{eqn:MEevol}).  We will use the notation $\phi(\mathbf{x},t)=\phi$ where this is unambiguous.

The `classical solution' to Eq. (\ref{eqn:Hunshifted}) is obtained by considering only terms at most bilinear in $\phi$ and $\tilde{\phi}$, corresponding to the noiseless case.  This has $\tilde{\phi}=1$, as is easily checked using the methods from Section \ref{subsec:sPDE}.  Therefore it is useful to perform a field shift $\tilde{\phi} \to \overline{\phi}+1$ to obtain a neater Hamiltonian:

\begin{equation}
H(\phi,\overline{\phi}) = D \int d^dx \left[-\overline{\phi} \bigtriangledown^2 \phi - \phi \overline{\phi} \bigtriangledown^2 \overline{\phi} -\frac{2}{\pmut }\overline{\phi}^2\phi \label{eqn:fullH}
\right]. 
\end{equation}
The variable we are working with here, $\phi$, does not correspond directly to the real density we measure, although the expectation value for the two is the same.  The density  \cite{Jannsen00FTManyFlavors} is $\tilde{\phi} \phi = \rho$, which can be obtained directly by defining $\tilde{\phi} =e^{\tilde{\rho}}$, and $\phi = \rho e^{-\tilde{\rho}}$; $\rho$ is a real valued field.  This allows us to obtain an explicit equation for the density; however, the exponentials must be considered as their sum expansion, and in our case they do not cancel.  We will later be able to show that the higher order terms are progressively less important when the population is large.  Writing the less important terms within sums, the Hamiltonian for the real density for evolution in a type space is therefore:

\begin{align}
H(\rho, \tilde{\rho}) = &\int d^dx \Big(-D\tilde{\rho}\bigtriangledown^2\rho - \frac{2D}{\pmut} \rho \tilde{\rho}^2 \nonumber\\
&- D\sum_{n=2}^{\infty}\frac{\tilde{\rho}^{n}}{n!}\bigtriangledown^2\rho -\frac{4 D}{\pmut} \sum_{m=2}^{\infty} \frac{\tilde{\rho}^{2m}}{(2m)!}\rho \Big). \label{eqn:Hrho} 
\end{align}
We will use two different Hamiltonians in the following analysis, $H(\phi,\overline{\phi})$ for the complex field, and $H(\rho,\tilde{\rho})$ for the real density field.  Note that the over-line $\overline{\phi}$ notation refers to variables in the shifted field, not an average.  

\subsection{Noise in Field Theory} \label{subsec:noise}

An equation for the time development of the distribution of particles is obtained by taking the functional derivative (see e.g. \cite{LeBellacFT})\footnote{Note that this is the reverse of the standard method to obtain a field theory from a stochastic PDE representation \cite{Bausch76-RenormFieldTheory}, and is simple to do in practice.} of the Action $\mathcal{S}$ with respect to the complex field $\overline{\phi}$.  Conversely, the equation for $\overline{\phi}$ is obtained by functional derivation of the Action with respect to $\phi$, which often gives a pathological equation for $\overline{\phi}(t)$.  It is however possible to remove $\overline{\phi}$ from the stochastic PDE for $\phi$ when the Action is quadratic in $\overline{\phi}$, by linearising the Action in  $\overline{\phi}$.  To do this we introduce an auxiliary field $\eta$, which will correspond to a noise field.  To see why $\eta(\mathbf{x},t)$ is a noise field, consider a single point in the field.  Suppose $\eta$ is Gaussian, uncorrelated noise\footnote{Many authors prefer to incorporate the variance and correlations into the noise, defining correlators $\langle \eta(\mathbf{x},t),\eta(\mathbf{x'},t') \rangle$ that absorb \emph{all} noise terms and their cross-correlations, which is the appropriate form for further calculations.  For clarity, we instead keep the noises simple and retain the explicit magnitudes, but must be careful to combine the noise terms for calculations.} with unit variance such that $\langle \eta(\mathbf{x},t),\eta(\mathbf{x'},t') \rangle = \delta(\mathbf{x}-\mathbf{x'},t-t')$ and $p(\eta(\mathbf{x},t)) = e^{-\eta^2/2}/\sqrt{2 \pi}$, then the Fourier transform of this is:

\begin{equation}
\frac{1} {\sqrt{2 \pi}} \int_{-\infty}^{\infty}{e^{-\eta^2/2}e^{-i 2\pi k \eta} d \eta} = e^{-2\pi^2k^2}.
\end{equation}
Therefore, by writing $q=\sqrt{2}\pi k$:

\begin{equation} 
e^{-q^2}=\frac{1} {\sqrt{2 \pi}} \int_{-\infty}^{\infty}{e^{-\eta^2/2}e^{-i\sqrt{2}q\eta} d \eta}.\label{eqn:noiseft}
\end{equation}
We can equate $q$ with (a convenient form of) the field $\overline{\phi}$ at a particular point $(\mathbf{x},t)$.  For example the final term in Eq. (\ref{eqn:fullH}) is of the form $-a\overline{\phi}^2\phi$ (and H appears with an additional $-$ sign), so we can identify $q= \overline{\phi}\sqrt{-a \phi}$ and the term translates with noise to $\overline{\phi}\sqrt{2a\phi}\eta$, where $\eta$ is Gaussian uncorrelated noise.
In this sense, the field $\exp(-\overline{\phi}^2)$ represents the `integrated out' form of the noise.  

We proceed to calculate the linearised version of the noise for the above problems, with terms appearing in the formalism as $\exp(-H)$.  We can replace the form $q^2$ with $-i\sqrt{2}q \eta$ in $S$, which if $q$ is already in the form $\overline{\phi}^2$ will give us the result immediately, or we can rearrange the result by parts.  If $q^2$ is negative (i.e. the original term was negative in $H$), this will lead to a real noise term, and conversely an imaginary noise term if $q^2$ is positive.  Firstly, we rearrange one noise term found in Eq. (\ref{eqn:fullH}) using integration by parts into an appropriate form:

\begin{equation}
-D\overline{\phi}\phi \bigtriangledown^2\overline{\phi} = -(D/2) \overline{\phi}^2\bigtriangledown^2\phi + D \phi(\bigtriangledown\overline{\phi})^2. \label{eqn:noisedecomp}
\end{equation}
Noise terms normally cannot be decomposed without consideration of cross-correlations.  Explicit consideration of correlations is complicated in this case as the $\overline{\phi}$ term appears within a gradient operator in some terms, but not in others.
Fortunately, we can perform decomposition to real and imaginary parts; although care must be taken \cite{HowardTauberRealIm} to determine the relative importance of combined real and imaginary noises, we can separate the real components (i.e. negative terms in $H$) and imaginary components (i.e. positive terms in $H$) in Eq. (\ref{eqn:fullH}) using Eq. (\ref{eqn:noisedecomp}):

\begin{equation}
H(\phi,\overline{\phi}) = D \int d^dx \left[-\overline{\phi} \bigtriangledown^2 \phi +\phi(\bigtriangledown\overline{\phi})^2- \overline{\phi}^2\left(\frac{\bigtriangledown^2\phi}{2} +  \frac{2}{\pmut}\phi\right) \label{eqn:fullHforNoise}
\right]. 
\end{equation}
Since the noise terms appear as $\exp(-H)$, so are the opposite sign to how they appear in $H$, they transform as follows:

\begin{align}
-D\phi(\bigtriangledown\overline{\phi})^2 &\to i \sqrt{2D\phi}\bigtriangledown(\overline{\phi})\eta \nonumber\\
&\to -i \bigtriangledown(\eta\sqrt{2D\phi}) \overline{\phi} \label{eqn:diffnoise}\\
D \overline{\phi}^2 \left(\frac{\bigtriangledown^2\phi}{2} + \frac{2}{\pmut}\phi\right) 
        &\to \sqrt{D(\bigtriangledown^2\phi + \frac{4}{\pmut} \phi)}\overline{\phi}\eta. \label{eqn:funnynoiseterm}
\end{align}
Eq. (\ref{eqn:funnynoiseterm}) also appears in the equation for the real noise from Eq. (\ref{eqn:Hrho}), using the $n=2$ term from the sum; the translation to a noise field is only valid when the remaining sums are discarded as there would be correlations to consider with the higher order terms.  Additionally, the imaginary Eq. (\ref{eqn:diffnoise}) term is absent as the $\rho$ field is constructed to be strictly real.  Also, we will later need the linearised form for the $ \overline{\phi}^2\phi$ term:

\begin{equation}
\frac{2D}{\pmut} \phi\overline{\phi}^2 \to 2\sqrt{\frac{D\phi}{\pmut }} \overline{\phi}\eta. \label{eqn:sqrtnoise}
\end{equation}
The above fields can be simply described.  Eq. (\ref{eqn:diffnoise}) represents so called `diffusive' noise (in the imaginary axis, for our case) - it is conserved (the differential ensures that what goes in at one point comes out at the next) and decays with $\phi$ as $\sqrt{\phi}$ (recall $\langle \phi \rangle = \langle n(\mathbf{x},t)/N \rangle$).  Eq. (\ref{eqn:sqrtnoise}) describes `square-root' (in magnitude) multiplicative noise.  Because it is non-conservative, multiplicative noise in general can have dramatic effects on the behaviour of the system.  We call Eq. (\ref{eqn:funnynoiseterm}) `mutation noise', as it arises from movement only on a mutated birth event.  

Representation of mutation noise as a stochastic PDE must be done carefully, as the term inside the square root may become negative when the gradient is large and negative. This is not physical and is due to representing the evolution process as continuous field theory, then forcing the field theory into a stochastic PDE.  Consider the discrete representation of the term inside the square root in Eq. (\ref{eqn:funnynoiseterm}):

\begin{equation}
\bigtriangledown^2 \rho(\mathbf{x},t) + \frac{4}{\pmut} \rho(\mathbf{x},t) =  \rho(\mathbf{x}+1,t) +  \rho(\mathbf{x}-1,t) + \left(\frac{4}{\pmut}-2\right) \rho(\mathbf{x},t).
\end{equation}
This is \emph{positive definite} since $\pmut \le 1$. We therefore impose the extra constraint that $\rho(\mathbf{x}+1,t) +  \rho(\mathbf{x}-1,t) > \bigtriangledown^2 \rho(\mathbf{x},t) > -2 \rho(\mathbf{x},t)$ on mutation noise.  This can be achieved by using $\Theta(\bigtriangledown^2\rho(\mathbf{x},t)+2\rho(\mathbf{x},t))$, where $\Theta(y)$ is the Heaviside step function; $\Theta(y)=0$ for $y \in (-\infty,0)$ and $\Theta(y)=1$ for $y \in(0,\infty)$.  This ensures positivity but the convergence properties as $\Delta \mathbf{x} \to 0$ are not currently known.

\subsection{Dimensional Analysis} \label{subsec:dimanneutral}

In order to establish which terms are important for the `large scale' behaviour of the system, dimensional analysis can be used.  This involves considering the contribution of terms at different scales by assigning dimensions to the constants (called coupling constants) of each term, and ensuring that the equations are dimensionally consistent.  The system is then rescaled and the constants will change according to their dimensions.  We consider the `long wavelength' limit, so that distance scales are much longer than any lattice spacing and the `fine structure' is averaged out.  The fine structure of models will depend on details such as the definition of the lattice and the exact form of mutations (i.e. whether strictly nearest neighbour or with some short ranged distribution such as exponential).  Fine structure is lost in the dimensional rescaling, but many models have the same phenomological description, i.e. are described identically at the macroscopic scale of large wavelengths.  Thus the rescaling can result in significantly simpler models in which only the most important details are retained.  

The following results will hold in the asymptotic limit of large population $N$, and apply only to the description at large scales.  
For all finite $N$ there will be a clustering length scale.  Together with $\kappa$, there would be two length scales in the problem and the appropriate dimensions for the coupling constant cannot be uniquely determined, hence dimensional analysis cannot be applied.

As all terms in the Action $S$ given by Eq. (\ref{eqn:action}) appear in an exponential (in Eq. (\ref{eqn:obsexp})) they must be dimensionless.  We define a wavevector $\kappa = h^{-1}$ as our unit of measurement (with $h$ a small length scale). Each term in the Hamiltonian $H$ contains an integral of dimensions $\kappa^{-d}t^{-1}$ ($d$ is the dimension of space), hence each term must have a spatial dimension of $\kappa^{d}$ and time dimension $t$.  Scaling of space will extract the relative importance of the terms in $H$; time scaling must then be performed to ensure that the equation retains the time derivative term in the Action $S$ with the dominant term(s) from $H$.  The time scaling is not of interest in this case and we will not consider it further.

There is an arbitrary choice when defining the dimensions of $\phi$ and $\overline{\phi}$, provided that $[\phi\overline{\phi}] = \kappa^d$; however, there is a `natural' choice, meaning a choice in which the scaling dimensions of the terms is correct.  We will identify the natural choice for our case in Section \ref{subsec:sPDE}, but progress can still be made without assuming a particular dimensionality of $\phi$ as some terms are irrelevant in the limit $\kappa \to \infty$ regardless of the dimensional assignment.

We will be rescaling to large $\kappa$, 
and hence only the highest order terms in $\kappa$ are `relevant', as they will dominate the effective equation at large $\kappa$.
We consider $D$ dimensionless, but introduce a coupling constant on all terms that contains all dimensional components; this constant has magnitude $1$ in the original, unscaled system.  The derivative $\bigtriangledown$ has scaling dimension $[\bigtriangledown] = \kappa^{1}$.  
The dimension of the fields are $[\phi] = \kappa^{d-\epsilon}$ and $[\overline{\phi}] = \kappa^{\epsilon}$, where $\epsilon$ is the parameter that controls the relative dimensions of the field and must be in $[0,d]$.  


We proceed with an analysis of the dimensions of the coupling constants.  
Performing the full Renormalisation Group analysis \cite{TauberHowardLee05} would explicitly perform the rescaling, providing details of how the scaling occurs and giving the natural choice of $\epsilon$ as a by product.  We don't perform this analysis, but instead consider all possible values of $\epsilon$ for now, and use previous results to identify the correct choice.  The coupling constants introduced will be called $a_i$, where $i$ is just a label.  The terms of interest in $H(\phi,\overline{\phi})$ from Eq. (\ref{eqn:fullH}) are:

\begin{align}
&[a_1 \overline{\phi} \bigtriangledown^2\phi] = \kappa^{d+2} [a_1] = \kappa^{d} \nonumber\\
&\implies [a_1] = \kappa^{-2}, \label{eqn:dim1}
\end{align}
\begin{align}
&[a_2 \phi\overline{\phi}\bigtriangledown^2\overline{\phi}] = \kappa^{2+d+\epsilon} [a_2] = \kappa^{d} \nonumber\\
&\implies [a_2] = \kappa^{-2-\epsilon}. \label{eqn:dimdelsq}
\end{align}
\begin{align}
&[a_3 \overline{\phi}^2 \phi] = \kappa^{d+\epsilon}[a_3] = \kappa^{d} \nonumber\\
&\implies [a_3] = \kappa^{-\epsilon},\label{eqn:dimofphi}
\end{align}


Recalling that $\epsilon \in [0,d]$, hence $\kappa^{-\epsilon} \ge \kappa^{-d}$, we can conclude the following.  
If we assume $\epsilon = 0$, then Eq. (\ref{eqn:dimofphi}) dominates both Eq. (\ref{eqn:dim1}) and Eq. (\ref{eqn:dimdelsq}). If we instead assume $\epsilon > 0$, then Eq. (\ref{eqn:dim1}) dominates Eq. (\ref{eqn:dimdelsq}), although the importance of the remaining terms cannot be determined without knowledge of $\epsilon$.  Therefore the term $\phi\overline{\phi}\bigtriangledown^2\overline{\phi}$ from Eq. (\ref{eqn:dimdelsq}) can be discarded, and the Hamiltonian $H_0$ for infinite $N$ can be written:

\begin{equation}
H_{0}(\phi,\overline{\phi})=\int d^dx \left[-D\overline{\phi}\bigtriangledown^2\phi - \frac{2D}{\pmut}\overline{\phi}^2\phi\right].\label{eqn:Hinf}
\end{equation}
Some of these terms are also present in $H(\rho,\tilde{\rho})$, but there are additional terms that appear when considering the sums in Eq. (\ref{eqn:Hrho}), where we have the minimum summand variables $m=n=2$ in the following terms:


\begin{align}
&[b_m \tilde{\rho}^{2m} \rho] = \kappa^{d+2(m-1)\epsilon}[b_m] = \kappa^{d} \nonumber\\
& \implies [b_m] = \kappa^{-(2m-1)\epsilon}.  \label{eqn:dimexp2}
\end{align}
\begin{align}
&[c_n \tilde{\rho}^n\bigtriangledown^2 \rho] = \kappa^{2+d+(n-1)\epsilon}[c_n] = \kappa^{d} \nonumber\\
& \implies [b_n] = \kappa^{-2-(n-1)\epsilon}, \label{eqn:dimexp1}
\end{align}
Hence we cannot yet truncate the exponential as all terms \emph{may} be important, as they only scale negatively for certain values of $\epsilon$.  However, in the case $\epsilon>0$ then the real field $\rho$ rescales to the same equation as the complex field $\phi$, i.e. $H_0(\rho,\tilde{\rho})=H_{0}(\phi,\overline{\phi})$.  Eq. (\ref{eqn:Hinf}) then provides the correct description of evolution in the infinite population limit.  At large but finite $N$ the importance of terms will correspond to their scaling dimension and hence we can make various levels of approximation in order to capture these details.  The approximate Hamiltonian for the real density is obtained from Eq. (\ref{eqn:Hrho}) by considering the first order correction to Eq. (\ref{eqn:Hinf}), given by the $n=2$ term from Eq. (\ref{eqn:dimexp1}):

\begin{equation}
H_{1} = \int d^dx \left[-D\tilde{\rho}\bigtriangledown^2\rho - \frac{D}{2} \tilde{\rho}^2\left(\frac{\bigtriangledown^2\rho}{2} - \frac{2 D}{\pmut} \rho\right) \right].
\label{eqn:Happrox}
\end{equation}

\subsection{Stochastic PDEs} \label{subsec:sPDE}

A stochastic PDE can sometimes be obtained from the field theory by calculating the functional derivative of the Action $S$, as discussed in Section \ref{subsec:noise}.  This is possible when all terms can be linearised, and we have presented such Hamiltonians at various levels of approximation.  Other Hamiltonians that cannot be made bilinear in the fields, such as Eq. (\ref{eqn:Hrho}) will not yield a stochastic PDE and will not be considered here.  However, the available forms are the most important, consisting of: 1. Eq. (\ref{eqn:fullHforNoise}), the exact equation for the evolution of the complex field $\phi$.  2.  Eq. (\ref{eqn:Hinf}), valid for $\phi$ in the infinite population limit, and we will find also valid for the real density $\rho$.  3. Eq. (\ref{eqn:Happrox}), the first order correction at large but finite population for the real density $\rho$.

The complete Hamiltonian $H(\phi,\overline{\phi})$ from Eq. (\ref{eqn:fullH}) is rearranged to Eq. (\ref{eqn:fullHforNoise}), which transforms with noise via Equations (\ref{eqn:diffnoise}) and (\ref{eqn:funnynoiseterm}) to:

\begin{equation}
H(\phi,\overline{\phi},\eta)=\int dx \left[ -D \overline{\phi}\bigtriangledown^2\phi -\sqrt{D\bigtriangledown^2\phi+\frac{4D\phi}{\pmut}}\overline{\phi}\eta_1+i\overline{\phi}\bigtriangledown(\eta_2\sqrt{2D\phi})\right]. 
\end{equation}
The two noise fields $\eta_1(\mathbf{x},t)$ and $\eta_2(\mathbf{x^\prime},t^\prime)$ are uncorrelated with unit variance, and form the real and imaginary parts of the noise with the given magnitudes.  Therefore considering the full action $S$ and taking the functional derivative with respect to $\overline{\phi}$, we obtain the stochastic PDE for the complex field in the evolution case:

\begin{equation}
\frac{\partial\phi(\mathbf{x},t)}{\partial t} = D \bigtriangledown^2\phi +\sqrt{D\bigtriangledown^2\phi+\frac{4D\phi}{\pmut}}\eta_1+i \bigtriangledown(\eta_2\sqrt{2D\phi}). \label{eqn:SPDEevolfull}
\end{equation}
This equation is valid at arbitrary population size $N$, and is an exact representation in the sense that it captures the finite population size effects correctly ($N$ appears via the density $\phi(\mathbf{x},t)=n(\mathbf{x},t)/N$).  The only approximation involved is the use of continuous time and space, but the same `amount of individual' will be moved in a time unit as in the discrete case, with equal variance in both space and time.

Similarly, the best possible stochastic PDE for the real density in evolution is obtained from Eq. (\ref{eqn:Happrox}):

\begin{equation}
\frac{\partial \rho(\mathbf{x},t)}{\partial t} = D \bigtriangledown^2 \rho(\mathbf{x},t) + \sqrt{D\left(\bigtriangledown^2\rho(\mathbf{x},t) + \frac{4\rho(\mathbf{x},t)}{\pmut}\right)}\eta(\mathbf{x},t), \label{eqn:SPDEevolapprox}
\end{equation}
with the added constraint that $\bigtriangledown^2\rho(\mathbf{x},t) > -2\rho(\mathbf{x},t)$.  The `mutation noise' appears because our individuals only `move' when they reproduce, rather than diffusing throughout their lifetimes.  We have to introduce a cutoff in the gradient to ensure that the mutation noise remains real.  

Finally, we will show that only the square-root noise is `relevant' in the infinite population limit, both for the real density $\rho$ and the complex field $\phi$, so in this case, $\phi=\rho$.  The stochastic PDE obtained in this case from Eq. (\ref{eqn:Hinf}) is:

\begin{equation}
\frac{\partial \rho(\mathbf{x},t)}{\partial t} = D \bigtriangledown^2 \rho(\mathbf{x},t)  + \sqrt{\frac{4D\rho(\mathbf{x},t)}{\pmut}}\eta(\mathbf{x},t), \label{eqn:SPDEreal}
\end{equation}
which is the equation for a birth/death process in which individuals diffuse in \emph{real} space, given by Eq. (\ref{eqn:GWtree}).


We complete the analysis with the deduction of the dimensions of $\phi$ using $\epsilon$ and hence justify our claim that $d_c=2$ and hence that the real space birth/death process is equivalent in the infinite limit to the evolution process.

\begin{enumerate}
\item The full evolution equation for the complex field $\phi$, given by Eq. (\ref{eqn:fullH}), is dimensionally dominated by either `square-root' noise or diffusion depending on $d$, and therefore we can take the large wavelength, infinite population limit and obtain the Hamiltonian for this process given by Eq. (\ref{eqn:Hinf}), hence the stochastic PDE given by Eq. (\ref{eqn:SPDEreal}).
\item Eq. (\ref{eqn:SPDEreal}) is the same as Eq. (\ref{eqn:GWtree}) from Super-Brownian Motion in all dimensions.  Hence we can establish that $d_c=2$ as in Super-Brownian motion, and by combining Equations (\ref{eqn:dim1}) and (\ref{eqn:dimofphi}) for the dimensions of the deterministic diffusion term and the square-root noise term respectively, we find that $\epsilon =d$.
\item Therefore the real field described by Eq. (\ref{eqn:Hrho}) can also be described by Eq. (\ref{eqn:Hinf}) in the large population limit, as the truncation of the extra terms present in the real field is now justified, as each is dimensionally less important in $d \ge 1$ than the terms retained (from Equations (\ref{eqn:dimexp2}) and (\ref{eqn:dimexp1})).
\item Finally, by considering the terms that decrease less quickly with $\kappa$, the stochastic PDE representing the leading order (large but finite $N$) correction to Eq. (\ref{eqn:Hinf}) is given by Eq. (\ref{eqn:Happrox}).
\end{enumerate}

\subsection{Numerical simulation of the stochastic PDEs} \label{subsec:langevin}

We have obtained approximate stochastic PDEs (alternatively, called Langevin equations) for evolution (Equations (\ref{eqn:SPDEevolfull}), (\ref{eqn:SPDEevolapprox}) and (\ref{eqn:SPDEreal})) and argued that the mutation noise term reduces to simple $\sqrt{\phi}$ noise in the large $N$ limit.  We also have an exact equation for the complex field $\phi$, which can be related to the real density distribution.  This is done by noting that in operator notation, operators can be reordered by using the permutation relation so that it is `normal ordered' (i.e. all $\hat{a}^\dagger$ are to the left of all $\hat{a}$).  The average of a normal ordered operator is identical to the average of the same operator with the $\hat{a}^\dagger$ operators removed \cite{TauberHowardLee05}; that is, $\langle (\hat{a}^\dagger)^m f(\hat{a}) \rangle = \langle f(\hat{a}) \rangle$.  We can normal order operators by using the commutation relation, remove the $\hat{a}^\dagger$ operators and then take the continuous limit as before.  Therefore:

\begin{eqnarray}
&\langle \rho \rangle = \langle \hat{a}^\dagger\hat{a} \rangle = \langle \hat{a} \rangle = \langle \phi \rangle, \\
&\langle \rho^2 \rangle = \langle \hat{a}^\dagger\hat{a} \hat{a}^\dagger\hat{a} \rangle = \langle \hat{a}^\dagger(\hat{a} + \hat{a}^\dagger\hat{a} \hat{a} \rangle = \langle \phi \rangle  + \langle \phi^2 \rangle.
\end{eqnarray}
Continuation to arbitrary moments is straightforward; each only depends on previous moments.  Therefore, if we can calculate (or simulate from an exact equation) the moments of the complex field $\phi$, we can calculate the moments of the real density exactly.

The final problem is of iterating a stochastic partial differential equation, although this is not a trivial task.  
We use the `splitting' method presented in \cite{PechenikMultNoise99,Moro04-SDEnumerics,MoroSchurz06-SDEsol} to accurately numerically integrate the stochastic PDE, where possible.  This method guarantees to give the correct results for the square-root noise term, but other terms can cause problems; in particular, $\phi=0$ is supposed to be an absorbing state and it should be an attractor.  The simplest approach of using Gaussian noise multiplied by $dt$ directly changes $\phi=0$ from an absorbing state to an unstable steady state, with probability $0$ of finding it at any finite $dt$.  The numerical solution of the $\Delta_t \phi(\mathbf{x},t) = \sqrt{\frac{D}{2}\bigtriangledown^2\phi + \frac{4 \phi}{\pmut}} \eta(\mathbf{x},t)$ term does not seem have an algorithm for finite $dt$ in the literature and so we use ad-hoc truncation to zero as described in Section \ref{subsec:noise}, though it should be noted that pursuing numerical integration is dangerous in this case.

\subsubsection{Numerical results}

\begin{figure*}[ht]
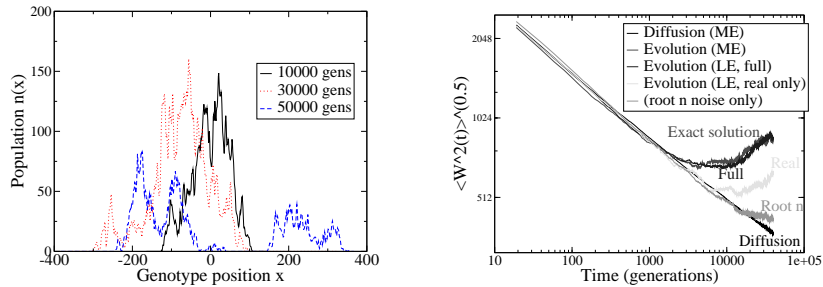

\begin{tabular}{cc}
  \begin{minipage}{.45\textwidth}
    \centering
    \epsfig{file=FigLangevinEvolDistrib.eps,width=0.9\textwidth}
  \end{minipage}
  &
  \begin{minipage}{.45\textwidth}
    \centering
    \epsfig{file=FigIWidth_Comparison_Analytical.eps,width=0.9\textwidth}
  \end{minipage}
\end{tabular}
\caption{Left: Evolved population distribution for the evolution case using the approximation of infinite N, given by Eq. (\ref{eqn:SPDEreal}).  Shown is the distribution at various times.  Note that it allows the distribution to split into multiple clusters as the original Master Equation, and local extinction is possible.  Qualitatively, the model appears accurate. Right: Interface width $\langle n(x)^2\rangle-\langle n(x)\rangle ^2$ as a function of time for the various relevant cases, with initial conditions of all population starting at position $0$.  Plotted are, from the bottom up, `1. Diffusion ME': the Master Equation evaluation of Diffusion (Eq. (7) from \cite{TauberHowardLee05}) which is indistinguishable on this plot from the diffusion stochastic PDE from Eq. (\ref{eqn:diffSPDE}). `2. Evolution (LE, root n noise only)': given by (Eq. (\ref{eqn:SPDEreal})). `3. Evolution (LE, real only)': the full solution given by Eq. (\ref{eqn:SPDEevolfull}) with the insistence that the field is real (truncation of negative densities to 0). `4. Evolution (LE, full)': the full solution (Eq. (\ref{eqn:SPDEevolfull})) itself.  Finally, `5. Evolution (ME)': the Master Equation evaluation of evolution (Eq. (\ref{eqn:MEevol})).  Only the full consideration of the complex solution yields the desired behaviour, although the qualitative dynamics of a peak are captured.  We use $D=0.25$ and $N(t=0)=10000$ throughout. \label{fig:LangApproxEvol}}
\end{figure*}

We now present the results of simulating the partial differential equations obtained for the diffusion case and the approximated evolution case, and show that they behave approximately as expected.  For comparison purposes, we also show the behaviour the stochastic PDE obtained from a simple diffusion of $N$ non-interacting particles (see Appendix \ref{appendix:diffusion}):

\begin{equation}
\frac{\partial \rho_D(\mathbf{x},t)}{\partial t} = D \bigtriangledown^2\rho_D(\mathbf{x},t) + \bigtriangledown(\eta(\mathbf{x},t)\sqrt{2D\rho_D(\mathbf{x},t)}). \label{eqn:diffSPDE}
\end{equation}
Note the presence of the $\bigtriangledown$ on the noise term, ensuring that this noise conserves particles locally.  Figure \ref{fig:LangApproxEvol} (Left) shows the general behaviour of the distributions for Eq. (\ref{eqn:SPDEreal}), containing only square root noise and corresponding to the `real space' case of reproducing and dying particles subject to spatial diffusion.  The Figure demonstrates that square root noise does produce clustering and permits local and global extinction.
However, we find that Eq. (\ref{eqn:SPDEreal}) only captures the qualitative aspects of the clustering.  Figure \ref{fig:LangApproxEvol} (Right) shows the (ensemble averaged) Interface Width\footnote{The Interface Width $\langle n(x)^2\rangle-\langle n(x)\rangle ^2$ is \emph{not} directly related to the standard deviation of the distribution, but rather the distribution is viewed as an interface.  The Interface Width describes the 'roughness' or deviation of the distribution from a straight line.} of the distribution against time, comparing the various cases.
We compare all of the stochastic PDEs we have encountered, and see that all terms are important quantitatively - the width is not correctly represented in any approximation, even at this fairly large value of $N=10000$.  The diffusion stochastic PDE (Eq. (\ref{eqn:diffSPDE})) fits the Master Equation solution (Eq. (7) from \cite{TauberHowardLee05}) closely.  Only diffusion and square root noise can be guaranteed to be accurate numerical integrations of the corresponding stochastic PDEs, due to the numerical problems discussed above.  The average interface width increases with time after passing some minimum value in all evolution cases, because the total population is not conserved.  Since we disregard runs in the ensemble average where the population becomes extinct the average population size tends to increase from its initial value\cite{Slade02-SuperBrownian}.

The full complex solution from Eq. (\ref{eqn:SPDEevolfull}) quantitatively captures the dynamics, although it must be admitted that the `ad-hoc' nature of the process for discretisation permits us to try different procedures and keep only those that worked.  This successful regime used a truncation of densities under a small amount chosen specifically for the time-step and total population, and the time-step was chosen very small.  Gaussian numbers were used for the generation of the noises.  

\section{Summary}


We found that death and reproduction with mutation in a type space is identically described in the large scale and population limit as diffusing particles undergoing birth/death processes, and is therefore described by a super Brownian Motion.  This tells us that the critical dimension for the evolution process is $2$ in Euclidean space.  Hence specialised models such as presented in \cite{Lawson06-Neutral} are essential for considering the distribution of a given phenotype ($d \leq 2$).  In higher dimensions $d>2$ lineage analysis is sufficient to describe the distribution of types developing in time, when coupled with the representation of type space.  However, all cases require a microscopic consideration of the underlying process for calculations at finite $N$.  Our simple Field Theory analysis has provided an exact description of the problem in the form of a stochastic PDE, Eq. (\ref{eqn:SPDEevolfull}).

We found the first order correction to the infinite $N$ stochastic PDE, given by Eq. (\ref{eqn:SPDEevolapprox}).  This is valid when the total population $N$ is large but finite.  Obtaining the correct stochastic PDE to represent a microprocess is non-trivial and mistakes are often made, as discussed in Ref. \cite{HowardTauberRealIm}, and hence a careful derivation such as ours is very important.  Our work brings together previous results and makes the underlying clustering process in evolution explicit.  Field theory is a tool that permits examination of finite systems and our work discusses the relevancy of stochastic PDEs in this case.


\section*{\small{Acknowledgements}}

\small{DJL acknowledges the EPSRC and the Scottish Government for funding, as well as Imperial College London where some of this research was carried out.  We are very thankful to Martin Howard and Uwe C. T\H{a}uber for invaluable discussions on the method.}

\appendix

\section{Calculating the Stochastic PDE for Diffusing particles}
\label{appendix:diffusion}

The starting point for diffusion is the Action for diffusing and non-interacting particles, obtained from Eq. 35 in \cite{TauberHowardLee05} for diffusing interacting particles, by setting the reaction rate $\lambda_0$ to zero:

\begin{equation}
A_{d}(\phi,\tilde{\phi}) = \int d^dx \int dt \left[\tilde{\phi}\partial_t \phi - D \tilde{\phi} \bigtriangledown^2 \phi \right].
\end{equation}
Here we have also neglected terms for initial and final conditions.  We first convert to a real density field $\rho$ using the methods from Section \ref{subsec:tofieldtheory}, by setting $\phi=\rho e^{-\tilde{\rho}}$ and $\tilde{\phi} - e^{\tilde{\rho}}$.  In this case the exponential terms cancel out and we are left, after integration by parts, with:

\begin{equation}
A_{d}(\rho,\tilde{\rho}) = \int d^dx \int dt \left[ \tilde{\rho}\partial_t \rho - D \tilde{\rho} \bigtriangledown^2 \rho + D \rho (\bigtriangledown\tilde{\rho})^2 \right].
\end{equation}
The final noise term is linearised using Eq. (\ref{eqn:diffnoise}) with but is of opposite sign, therefore giving the linearised Action:

\begin{equation}
A_{d}(\rho,\tilde{\rho}, \eta) = \int d^dx \int dt \left[ \tilde{\rho}\partial_t \rho - D \tilde{\rho} \bigtriangledown^2 \rho + \bigtriangledown(\eta\sqrt{2D\rho}) \tilde{\rho} \right],
\end{equation}
which on functional differentiation with respect to $\tilde{\rho}$ yields Eq. (\ref{eqn:diffSPDE}).

\end{document}